\documentstyle[11pt,emulateapj]{article}

\lefthead{Shuping et al.}
\righthead{CO and C$_2$ Absorption Toward W 40}
\begin{document}

\newcommand{\ctwo}{C$_2$}
\newcommand{\av}{$A_V$}
\newcommand{\wforty}{W40 IRS 1a}
\newcommand{\twelveco}{$^{12}$CO}
\newcommand{\thirteenco}{$^{13}$CO}

%for emulate apj
\submitted{Preprint -- To appear in The Astrophysical Journal (Accepted 9
February 1999)}

\title{CO and C$_2$ Absorption Toward W40 IRS 1a}

\vspace{10pt}
\author{R. Young Shuping\altaffilmark{1} and Theodore P. Snow\altaffilmark{1,2}}
\affil{Center for Astrophysics and Space Astronomy \\ University of Colorado, Boulder, Co. 
       80309-0389 \\ shuping@casa.colorado.edu \\ tsnow@casa.colorado.edu}

\vspace{10pt}

\author{Richard Crutcher\altaffilmark{2}}
\affil{Department of Astronomy \\ University of Illinois, 205 Astronomy Bldg., 1002 W. Green St.,
Urbana, Il. 61801 \\ crutcher@astro.uiuc.edu}

\and

\author{Barry L. Lutz\altaffilmark{3}}
\affil{Department of Physics and Astronomy \\ Northern Arizona University, Box 6010, Flagstaff, Az. 
       86011-6010 \\ barry.lutz@nau.edu}

\altaffiltext{1}{Visiting Astronomer, United Kingdom Infrared Telescope (UKIRT), operated by the Joint
	Astronomy Centre (JAC) for the Particle Physics and Astronomy Research Council (PPARC).}
\altaffiltext{2}{Visiting Astronomer, Infrared Telescope Facility (IRTF), operated by the University of 
	Hawai'i for the National Aeronautical and Space Administration (NASA).}
\altaffiltext{3}{Visiting Astronomer, Kitt Peak National Observatory (KPNO), operated by the Association
	of Universities for Research in Astronomy (AURA), Inc. under cooperative agreement with the
	National Science Foundation (NSF).}

\begin{abstract}
The H II region W40 harbors a small group of young, hot stars behind roughly 9
magnitudes of visual extinction. We have
detected gaseous carbon monoxide (CO) and diatomic carbon (C$_2$) in absorption
toward the star W 40 IRS 1a.  The 2-0 R0, R1, and R2 lines of
$^{12}$CO at 2.3
$\mu$m were measured using the CSHELL on the NASA IR Telescope Facility
(with upper limits placed on R3, R4, and R5) yielding an
$N_{CO}$ of $(1.1 \pm 0.2) \times 10^{18}$ cm$^{-2}$.  
Excitation analysis indicates $T_{kin} > 7$ K. The
Phillips system of C$_2$ transitions near 8775 \AA\ was measured using the Kitt Peak
4-m telescope and echelle spectrometer.  Radiative pumping models indicate a
total  C$_2$ column density of $(7.0 \pm 0.4)
\times 10^{14}$ cm$^{-2}$, two excitation temperatures (39 and 126 K), and a total gas density
of $n \sim 250$ cm$^{-3}$.  The CO ice band at 4.7 \micron\ was not detected, placing an upper
limit on the CO depletion of $\delta < 1\%$.  We postulate that the sightline
has multiple translucent components and is associated with the W40 molecular cloud.
Our data for
\wforty\, coupled with other sightlines, shows that the ratio of CO/\ctwo\ increases
from diffuse through translucent environs.  Finally, we show that the hydrogen to dust ratio
seems to remain constant from diffuse to dense environments, while the CO to dust ratio
apparently does not.

\end{abstract}

%\keywords{}

\section{Introduction}

Molecular species are useful diagnostics for understanding the physics and chemistry of the
interstellar medium (ISM), and have been studied in various ways for many years.  Most
molecules are observed in dense molecular clouds via rotational emission lines in the radio
band.  These emission studies are indispensable to our understanding of galactic ecology and the
ISM:  They yield maps of dense regions, have very high spectral (velocity) resolution, and allow
us to study complex molecules not otherwise observable.

There are some drawbacks to molecular emission studies, however.  First, analysis of the
line excitation is very model-dependent and can lead to significant systematic errors. Second, since
spatial resolution is typically low (and varies from species to species), it is sometimes hard
to discern whether emission from different species comes from the same location within a cloud.
This makes comparisons among species difficult.

Molecular absorption line observations have some important advantages over emission line
studies.  First, absorption from different species observed in the same line of sight are more
likely to coexist spatially, thus reducing geometrical ambiguities and allowing more reliable
inter-species comparisons.  Second, because the unexcited molecules are observed directly, column
density determinations are not as model-dependent as for emission studies, and hence more
accurate.  Third, some molecular species (notably H$_2$ and \ctwo) are essentially unobservable in the
radio due to lack of a dipole moment.  These species can sometimes be observed via electronic
transitions in the ultraviolet (e.g., \cite{sb82}), or through rotational-vibrational
transitions in the visible or infrared (e.g.
\cite{fetal94}; \cite{letal94}).  And finally, molecular abundances derived from absorption
measures can easily be compared to line-of-sight bulk properties (e.g., extinction,
polarization, and atomic abundances) which are also derived from absorption measures.  

CO and \ctwo\ have transitions at 2.3 $\mu$m and 8775 \AA, respectively, which can be
observed in absorption.  The primary drawback to studying molecular absorption lines in the infrared
(IR) is finding a suitable background star.  The target must be bright in
the IR yet fortuitously placed behind a substantial amount of absorbing material.  

In our effort to find background sources for absorption line studies in regions also
accessible to emission-line studies, we have selected the target \wforty, an OB star embedded
within the radio source W40. This source appears well-suited for our purposes; Though dim in the visible ($V
= 15.0$) it is bright in the IR ($K = 5.6$), and lies behind about 9 magnitudes of extinction  in $V$ (\av).
W40 itself is a blister-type \ion{H}{2} region breaking out of a local molecular cloud, about
400 pc away towards the galactic center (\cite{zl78}; \cite{cc82}).  A group of hot stars (most
likely B-class) is bathing the region with ionizing radiation; W40 IRS 2a (OS 2a) seems to be
producing most of the energy (\cite{setal85}).
IRS 1a is brighter than 2a in the optical and near-IR indicating somewhat greater extinction
toward IRS 2a.  Crutcher
and Chu (1982) mapped $^{12}$CO and $^{13}$CO, HCO$^+$, HCN and H$\alpha$ emission toward W40,
generating a comprehensive kinematic interpretation of the region.  Vall\'{e}e et al. 
(1987, 1991, 1992, and 1994) have
done extensive studies on the properties of the molecular cloud and its interaction with the
\ion{H}{2} region using \ion{C}{2} recombination lines, radio continuum emission, and CO
emission.    The existence of circumstellar dust shells around W40 IRS 1a, 2a and 3a ($T \sim
250-350$ K and
$M < 0.1 M_\odot$) has been suggested based on broad-band IR imaging and continuum
measurements in the millimeter and sub-millimeter (\cite{setal85}; \cite{vm94}).  

Both CO and \ctwo\ are important interstellar molecules.  
CO is the most abundant molecule in the ISM after
molecular hydrogen, and has a number of important roles:  
The CO emission lines at 2.6 mm and shorter are not
only responsible for cooling molecular clouds but also serve as tracers for H$_2$ and dense
molecular gas (e.g., \cite{mo95} and references therein).  \ctwo\ is much less abundant than CO
and has been observed primarily in diffuse and translucent clouds (e.g.
\cite{vdb89}; \cite{lsf95}).  Like H$_2$, C$_2$ has no dipole moment and hence no pure rotational
spectrum. It is a very useful diagnostic of cloud physical conditions such as kinetic
temperature, density, and radiation field intensity (\cite{vdb82}; \cite{vd84}).  Both CO and
\ctwo\ are important to carbon chemistry everywhere in the ISM.

CO absorption at 2.3 $\mu$m ($v = 2 - 0$) has not been observed as frequently as the v = 1 - 0
band at 4.7 $\mu$m, most likely owing to the smaller transition probabilities.
It is, however, easier to observe at 2.3 $\mu$m since
the thermal background from the Earth's atmosphere is much lower than at 4.7 $\mu$m.  Black and
Willner (1984) and Black et al. (1990) used the lines at 2.3 $\mu$m to study the physical conditions
and chemistry toward NGC 2024, NGC 2264 and AFGL 2591.  Recently, Lacy et al. (1994) were able to
observe both CO and H$_2$ absorption near 2.3 $\mu$m toward NGC 2024 IRS 2 yielding for the first time
a direct measure of the H$_2$/CO ratio in a molecular cloud.

Absorption lines of \ctwo\ have been used to address a number of problems in the ISM, including: 
Diffuse interstellar cloud chemistry (e.g. \cite{lsf95}), carbon chemistry in translucent clouds
(e.g. \cite{vdb89}), molecular cloud envelopes (e.g. \cite{fetal94}), and the line of sight
structure toward $\zeta$ Oph (e.g. \cite{crawford97}).  Federman et al. (1994) provide a good
compilation of \ctwo\ measurements up to 1992.

In this paper we report on absorption-line studies of CO and \ctwo\ toward \wforty.  The line of
sight is discussed in the next section.  In Section 3 we describe our observations and the
results, and in Section 4 we discuss their implication for the physical state of
the material in this line of sight.  The final section contains a brief summary of our
conclusions.

\section{Line Of Sight}

Very little is known about the interstellar sightline to \wforty.  Both Crutcher and Chu (1982)
and  Smith et al. (1985) infer $A_V \sim 9$.  The average extinction due to diffuse
material over 400 pc is about 0.6 mag (see \cite{s78}, p. 155).  Therefore the obscuring
material is most likely local to the W40 region and more dense than the diffuse ISM. In
addition, Crutcher and Chu (1982) found a molecular \thirteenco\ component in front of the W40
\ion{H}{2} region  with $V_{LSR} = 8$ km s$^{-1}$, $N_{CO} \simeq 1.2 \times 10^{18}$
cm$^{-2}$, and
$\log{(nT)}
\sim 4$, further suggesting that the sightline passes through cold material, most likely
associated with the neighboring molecular cloud.

The line of sight intersects at least 3 distinct physical regimes.  As noted above there
appears to be some cold foreground material, perhaps associated with the nearby
molecular cloud.  Closer to \wforty\, there must be a photon-dominated region (PDR), and
closer still, the W40 \ion{H}{2} region itself.  If \wforty\ has a circumstellar dust shell,
it is not clear what its effect on the CO and \ctwo\ spectra might be.  A warm dust shell might be
associated with an elevated gas temperature, and hence the appearance of high-$J$ molecular
rotational lines.  

The \ion{H}{2} region is roughly 0.9 pc across and has an electron density of 200
cm$^{-3}$ (based on results in Crutcher and Chu [1982] and a distance of 400 pc). The column
density of ionized hydrogen associated with the \ion{H}{2} region should be about $3 \times
10^{20}$ cm$^{-2}$, assuming spherical geometry and that all hydrogen is ionized.  Using the
relation of total hydrogen column density to extinction found in Bohlin, Savage, \&
Drake (1978), the \ion{H}{2} region should contribute
$A_V \sim 0.1$ mag to the line-of-sight extinction.  This, of course, assumes a standard
gas-to-dust ratio, which may not apply to  \ion{H}{2} regions in general.  The
contribution to
\av\ could increase or decrease depending primarily on the nature of any grain-destroying shocks
which may have passed through the region.  In either case, the 
\ion{H}{2} region probably does not strongly contribute to the total extinction on the line of sight
to \wforty.  In addition, molecules like CO and \ctwo\ cannot survive in the harsh  \ion{H}{2}
environment, so the  \ion{H}{2} region should not contribute to the line-absorption for
these molecules either.

There is almost surely a photon-dominated region (PDR) on the line of sight to \wforty, which
could account for as much as $A_V \sim 10$, nearly the entire measured visual extinction (see
\cite{hollenbach90} for a good discussion).  Hydrogen is expected to be neutral or molecular
throughout the PDR.  CO and perhaps \ctwo\ cannot survive in the regions of the PDR closest to
the 
\ion{H}{2} region.  The effect of dust in PDRs is poorly understood, and so a quantitative treatment
of these regions in general is difficult.

In summary, CO and \ctwo\ absorption lines should sample the cold material in the
foreground, at least part of the PDR, and none of the  \ion{H}{2} region.  A circumstellar
dust shell, if it exists, would most likely only contribute to high-$J$ molecular absorption.  The
visual extinction ($A_V
\sim 9$) should arise almost entirely in the foreground material (which may be part of the local
molecular cloud) and the PDR, though we note that the PDR could in theory produce all of the
observed extinction.

\section{Data and Analysis for \twelveco\ and \ctwo\ toward \wforty}

\subsection{C$_2$ Observations at 8775 \AA}

The observations of \wforty\ were obtained with the Cassegrain echelle
spectrograph and the RCA CCD camera on the 4-m Mayall telescope at Kitt
Peak National Observatory on the night of UT 16 June 1983. Two spectra,
each of one hour's duration, were recorded in the region of the 2-0 band of
the Phillips system (A$^1 \Pi_u$ -- X$^1 \Sigma_g^+$) of \ctwo.  The entire 2-0 band was
contained in a single order with its center near 8775 \AA\ and with a nominal
reciprocal dispersion of 0.074 \AA\ per pixel at the face of the CCD chip.
Quartz lamp and thorium-argon spectra were obtained for flat-fielding and
wavelength calibration.  An 84 $\mu$m wide slit was employed, providing
a nominal resolution of 0.15 \AA\ which in turn corresponds to approximately 2
pixels on the chip.

The two consecutive one-hour frames were co-added and averaged. Similarly,
two separate flat-field frames were co-added and averaged, and the result
was divided into the averaged stellar frame to produce the final
photometric spectrum.  Both flat-field and stellar frames were
bias-corrected.  The average of fifteen bias frames was first subtracted
from the flat-field and stellar frames, after which a second-order bias
correction was accomplished by subtracting from the average stellar and
flat-field frames the mean row biases obtained from the masked bias columns
of each.

The final spectrum was extracted by collapsing the three columns along the
slit direction which contained the maximum signal. This spectrum is shown
in Figure 1.  Spikes are due to cosmic ray hits and possibly OH airglow lines.
Rotational lines in all three branches (P, Q and R) were identified, with
the rotational quantum number $J$ reaching as high as 12 for the Q-branch.
Unfortunately, there were problems with the wavelength calibration at the telescope and we 
were not able to derive a precise measure for the radial velocities of the \ctwo\ lines.
Equivalent widths for these lines were determined using standard spectral
reduction procedures in the NOAO/IRAF package, and column densities for
each of the rotational levels were calculated from the rotational lines
using a simple Gaussian curve-of-growth (c.f. \cite{s78}).  
Initially, the $f$-value derived by Erman et al. (1982) was used.  Recently, the Phillips system
transition probabilities have been refined (see \cite{lsf95} and references therein) and the column
densities have been adjusted to reflect the new $f$-value derived by Lambert, Sheffer, and Federman
(1995), $(1.23 \pm 0.16) \times 10^{-3}$. The data for each line and the column densities for each
$J$ are shown in Tables 1 and 2.  Several of the
lines exhibited effects of saturation, and an estimate of the Doppler constant ($b$) was derived
by requiring that rotational lines originating from the same rotational energy level yield the
same rotational population for that level.  This method is a simple extension of the doublet
ratio method used in the analysis of saturated atomic species.  The best value for the Doppler
constant was found to be 1.25 km s$^{-1}$.  The $J = 12$ data are not included due
to high uncertainty:  Only the Q12 line was observed and it is very weak.

\subsection{\twelveco\ Observations at 2.3 $\mu$m}

The $v=2-0$ rotational-vibrational lines for CO fall near 2.3 $\mu$m, within the K band. 
Since the oscillator strengths are smaller, the $v=2-0$ lines are not as saturated as
their
$v=1-0$ cousins at 4.7 $\mu$m.  In addition, it is somewhat easier to observe at 2.3
$\mu$m, as the sky emission at 4.7 $\mu$m is much greater and more problematic. 
Contamination from stellar CO absorption should be negligible
since \wforty\ appears to be a hot O or B star, which would not allow stellar CO to
survive (\cite{cc82}; \cite{setal85}).  Circumstellar gas and dust, if it exists, could affect the high-$J$
molecular levels.

Ro-vib transitions at 2.3 $\mu$m for \twelveco\ were observed UT 11 June 1994 and
UT 21 July 1997 using the CSHELL IR spectrometer at NASAs Infrared Telescope Facility (IRTF) on
Mauna Kea. The IRTF is a 3 meter primary, off-axis cassegrain yielding f/13.67 at the
spectrometer slit.  The CSHELL is a cryogenically cooled echelle spectrometer with a 256 x 256
SBRC InSb detector array (\cite{greene93}).   We used the 0.5 \arcsec\ slit which gives $R
\simeq 43,000$.    Only the R0, 1, and 2 transitions were detected (Figure 3) with upper
limits placed on R3, 4 and 5.   A summary of the observations is shown in Table 3.  The data
were obtained and reduced following typical IR observing procedures which we briefly summarize
below.

Dark frames were coadded and subtracted from the flat-field image for
each wavelength setting.  \wforty\ and the standard stars were observed in both the
``A'' and ``B'' beams (each beam places the spectrum in a different spatial location on the
detector).  Beam differences (A-B and B-A) were calculated to eliminate sky
emission, then coadded and normalized using the flat field for the appropriate wavelength
setting.  Wavelength calibration was achieved using Ar, Kr, and Xe lamps with 3 lines per
wavelength setting.  One dimensional spectra for \wforty, the standard stars, and the
calibration lamps were then extracted from the detector images using the APALL task in
IRAF.

Once the wavelength solution was applied, telluric absorption features were identified in
both the standard stars and \wforty\ spectra.  Telluric lines were
eliminated from the \wforty\ spectra using the IRAF task TELLURIC, which shifts and
scales the standard star spectrum before dividing into the object spectrum.  It is important to
note that merely dividing the standard star spectrum into the object spectrum does not produce
the {\em true} object spectrum without telluric features.  To properly remove telluric features,
one must generate an {\em expected} object spectrum, convolve it with the standard star
spectrum, and compare to the actual observed object spectrum.  
Lacy et al. (1994) give a good discussion of this technique. We opted to merely divide,
however, as our data quality did not warrant more sophisticated techniques. 

The R0, 1, and 2 lines are
shown in Figure 3.  These lines imply $v_{lsr} = 2 \pm 2$ km s$^{-1}$, in contrast with the 8 km
s$^{-1}$ foreground molecular material seen in emission (\cite{cc82}).  Since emission line data
sample a large, beam-averaged area, and absorption lines just a pencil beam, we do not
necessarily expect velocities derived from both to agree.  We merely note that they are not
wildly different.

$R = 43,000$ corresponds to a resolution of $\Delta v \simeq 7$ km s$^{-1}$.  The
R0, 1, and 2 lines all have FWHM $\sim 8$ km s$^{-1}$ and hence are not well-resolved.  After
continuum normalization, each line was directly integrated for equivalent width.  Errors
reflect continuum placement ambiguity and are of high confidence ($2\sigma$).  All the widths
(along with transition information and errors) are given in Table 4.  Widths for the R3, R4 and
R5 lines are 2$\sigma$ upper limits based on the noise in the continuum at the expected line
position.  

The equivalent width of any absorption line is dependent on the column density of the species
and, if saturated, the velocity parameter {\em b} (c.f. \cite{s78}).  We generated a model curve
of growth (COG) using $b = 1.25$ km s$^{-1}$, as
determined from the C$_2$ lines. Each \twelveco\ equivalent width was fit to
the COG independently and the column density for each rotational level ($N_J$) is given in
Table 5.  The R2 through R5 lines were clearly optically thin while the R0 and R1
lines were more optically thick, falling on the transition from the linear portion to the
saturated part of the COG. If the $b$-value for CO is greater than 1.25 km s$^{-1}$, then the
R0 and R1 lines become more optically thin and their abundances drop by 2 and 0.5 $\times
10^{17}$ cm$^{-2}$ respectively. Summing the abundances for each line gives a total column
density of
$N_{CO} = (1.1 \pm 0.2) \times 10^{18}$ cm$^{-2}$ (assuming $^{12}$CO to be the dominant
isotope).  Depending on the excitation of the higher $J$ levels, this value could be slightly
too small, and if $b > 1.25$ km s$^{-1}$, then it would be too high.  Our column density for CO
is nearly identical to that inferred by Crutcher \& Chu (1982) for the foreground, $N_{CO}
\simeq 1.2 \times 10^{18}$ cm$^{-2}$.

\subsection{CO Ice Band at 4.7 \micron}

In an effort to detect the CO ice band at 4.7 \micron\ we made observations with moderate
spectral resolution at the United Kingdom IR Telescope (UKIRT) using the CGS4 on 19 August
1998.  All observations were carried out while nodding along the slit to remove atmospheric
emission.  The total integration time on \wforty\ was 11.2 minutes at an average airmass of
1.45.  The spectrum for
\wforty\ was ratioed by BS 7236 (B9V) to remove telluric absorption features.  The S/N is 
$\sim 44$ at 4.67 \micron\ and no CO ice feature is apparent.  An upper limit for the optical
depth of the band is
$\tau_{4.67} < 0.02$ (2$\sigma$), implying $N_{CO}(Ice) < 10^{16}$ cm$^{-2}$
(\cite{ta87}).  The depletion of CO into icy mantles for the \wforty\ sightline must be less
than 1 \%.

\section{Discussion}

\subsection{Excitation Conditions and the Foreground Cloud Physical Properties}

We have three basic diagnostics for the temperature and density toward \wforty.  The $^{13}$CO
emission data imply $\log{(nT)} \sim 4$, much lower than the nearby molecular
cloud, $\log{(nT)} \sim 6.5$  (\cite{cc82}).  Excitation conditions can also be derived from the
rotational populations of CO and \ctwo.

The rotational level populations of the \ctwo\ Phillips system ($v
= 0,J$) are attained via radiative pumping. Lifetimes of these levels are so long that
collisions and upward electronic transitions are the most important depopulation mechanisms.
Hence the rotational populations reflect the competition between pumping and collisions and are
non-thermal in general (\cite{vdb82}). For densities greater than $100$ cm$^{-3}$ the $J = 0$
and $J = 2$ levels are very nearly thermal.  Otherwise, the populations depend on the thermal
temperature, $T$, and the radiation parameter, $\frac{n_c \sigma}{I_R}$, where $n_c$ is the
collision partner density (usually $n($H$) + n($H$_2)$), $\sigma$ is the effective cross section
for collisional de-excitation, and
$I_R$ is the scaling factor for the radiation field in the far-red (\cite{vdb82}).  

Total molecular abundance and excitation temperatures were calculated from
models which fit a Maxwell-Boltzmann population distribution to the
observed rotational level abundances. As had been found for other
relatively dense clouds (\cite{lc83}), the rotational
abundances could not be fit with a single temperature distribution,
presumably as a result of radiative pumping (\cite{chaffee80}; \cite{vdb82}).
Consequently, we fit these data with a two-temperature model used by Lutz
and Crutcher (1983): J-levels 4 through 10 were fit with an excitation
temperature ($T_{ex}$) of 126 K, which we associate with the effects of radiative
pumping. After correcting the populations in J = 0 and 2 for the
contributions from the 126 K population distribution, we derived a J=2/J=0
excitation temperature of 39 K, which we associate with the thermal
distribution of the gas.  The resulting total column density of \ctwo\ towards
\wforty\ is $(7.0 \pm 0.4) \times 10^{14}$ cm$^{-2}$, based on a Boltzmann distribution for both
temperatures.  This is the highest column density of
\ctwo\ yet seen in absorption.  Figure 2 shows the final fits.

In comparing the results for \wforty\ to the radiative pumping models
calculated by van Dishoeck (1984), we found that they are best
characterized by her model with a kinetic temperature of 40 K and a
radiation parameter, $\frac{n_c \sigma}{I_R}$ of $5.27 \times 10^{-14}$.  Assuming $I_R = 1$,
$\sigma \sim 2 \times 10^{-16}$ cm$^2$ (\cite{vdb89}), and that hydrogen (\ion{H}{1} + H$_2$) is
the only important collision partner, we get $n_H \sim 250$ cm$^{-3}$ for the \wforty\
sightline.  Since the value of $\sigma$ is not known to better than a factor of 2
(\cite{vdb89}), our value for $n_H$ is equally imprecise.  
In addition, for an enhanced (or depleted) radiation field, the estimated
collision (hydrogen) density would scale accordingly.

Rotational transitions of CO are primarily excited by collisions with hydrogen. 
De-excitation can occur via collisions with hydrogen or by spontaneous line emission.  The
critical density at which collisions begin to overtake spontaneous emission is around 3000
cm$^{-3}$.  More detailed studies of the excitation, photodissociation, and chemistry have been
conducted by van Dishoeck \& Black (1988) and Warin, Benayoun, \& Viala (1996).  In general
it is found that the rotational populations of CO are sub-thermal except for the first few
levels in dense cases (\cite{wbv96}).  Hence the assumption of LTE will almost always produce
excitation temperatures which underestimate the actual thermal temperature ($T_{ex} <
T_{kin}$).  Recent work by Wannier, Penprase, \& Andersson (1997) suggests that the dominant
form of CO excitation in diffuse and translucent clouds can be line emission from nearby
molecular clouds, if the clouds have similar velocity vectors.  

As a start, we constructed a Boltzmann plot for the \twelveco\ data (Figure 4).
Fitting the temperature to all lines gives $T_{ex} = 7$ K.  This value is similar to that found by Cruther \&
Chu (1982), but it is important to note that the excitation temperatures for \thirteenco\ and \twelveco\ are
different in general (\cite{wbv96}).  Since the thermal temperature is greater than the CO rotational
temperature in general, 
$T_{kin} > 7$ K, which is consistent with the \ctwo\ analysis.  Warin, Benayoun, \& Viala (1996) have
constructed models of CO excitation for the dense, translucent, and diffuse cloud regimes, including
UV photodissociation.  These models show very clearly that the excitation temperature is much
lower than the thermal temperature for diffuse and translucent clouds. 
The excitation of CO in dense clouds is nearly thermal and $T_{ex} = T_{kin}$. 
If we assume that the cloud(s) toward \wforty\ are dense, then $T_{kin} \simeq 7$ K.
For diffuse and translucent clouds, the level populations observed can be scaled (see Section
4.2 of Warin, Benayoun, \& Viala [1996]) to derive ``pseudo-LTE'' rotational
populations, i.e. representative LTE populations for the actual thermal temperature of the gas. 
A slope can be fitted to these populations and a thermal temperature inferred.
If the cloud(s) toward \wforty\ are translucent, then the R0 population is reduced by
$\sim 1.75$ (changes in R1 and R2 happen to be negligible).  The resulting populations imply
$T_{kin} \simeq 9$ K (see Figure 4), which still does not agree with the \ctwo\ analysis.  It is
not clear, however, that the models constructed by Warin, Benayoun, \& Viala (1996) apply to
our line of sight:  The UV field is probably higher than normal near W40, and it seems very
likely that there are multiple clouds on the sightline.

Since the cloud(s) on the \wforty\ line of sight are very near the W40 molecular cloud, and
the LSR velocities of both are similar (5 and 8 km s$^{-1}$), the excitation for
\twelveco\ may be at least partially radiative.  Comparing to models constructed by Wannier,
Penprase, \& Andersson (1997) it is apparent that our $T_{ex} = 7$ K is degenerate:  It can be
accounted for by many combinations of radiative excitation, collision partner density, and kinetic
temperature.  The \ctwo\ excitation conditions can help constrain those for \twelveco.  If we assume $T_{kin}
\simeq 40$ K, the density of the gas can range from 0 to 450 cm$^{-3}$, depending on the efficiency of
radiative excitation.  Collisional excitation becomes dominant as $n$ approaches 450 cm$^{-3}$ (which we use
as an upper limit).  If we further assume $n \sim 250$ cm$^{-3}$ (as indicated by the \ctwo\ data),  then the
excitation of CO can only be explained by collisional and radiative processes combined.  

A summary of the \wforty\ sightline cloud physical properties based on the CO and \ctwo\ data
in this paper as well as the CO emission study by Crutcher and Chu (1982) is given in Table 6. 
As discussed in Section 2, we assume that nearly all of the absorption on this line of sight
arises in one cloud complex (single or multiple components) local to the W40 region, since it is
only 400 pc distant (i.e., the cloud(s) cannot be diffuse).  The physical conditions are best determined by
\ctwo: $n
\sim 250$ cm$^{-3}$ and $T \simeq 40$ K.  The CO emission and absorption values and limits agree very
well.  The cloud(s) are clearly not dense enough to be considered molecular, but could be considered
translucent.  Translucent lines of sight typically have $A_V = 2 -- 5$ and can be studied via both absorption
lines (UV, optical, and/or IR), and radio emission lines.  In addition, translucent material is expected
in the outer envelopes of molecular clouds (\cite{vdb89}).  We
postulate that the \wforty\ sightline is composed of multiple translucent components associated with the W40
molecular cloud.

\subsection{Molecular Abundances}

CO and \ctwo\ abundances have been determined jointly for a number of sightlines, a sample of which is shown
in Table 7.  
\wforty\ is one of few sightlines allowing a direct comparison of CO and \ctwo\ in
{\em absorption}.  The abundance ratio we derive is CO/\ctwo\ $= 1600 \pm 600$ (2$\sigma$). 
The depletion of CO onto dust is negligible ($< 1 \%$), in view of our failure to detect CO ice, but the \ctwo\
depletion has not been assessed.  Note that
the CO/\ctwo\ ratio increases from diffuse to translucent and molecular regimes
indicating (to first order) that the formation/destruction rates favor CO over \ctwo\
as cloud type changes.  This may merely be due to self-shielding of CO, but other processes may
also be in action:  The data are not yet precise enough to tell.

Assuming all hydrogen is molecular, we can estimate the
amount of H$_2$ on the \wforty\ line of sight from
$N_{CO}$.  Using an H$_2$/CO ratio of $3700_{-2700}^{+3100}$
(based on the direct comparison of H$_2$ and CO IR absorption lines
toward NGC 2024 IRS 2 ($A_V = 21.5 \pm 5$), \cite{letal94}), we find $N_{H_2} =
4.4_{-3.2}^{+3.8}
\times 10^{21}$ cm$^{-2}$.  The  H$_2$:CO ratio is derived from a
direct comparison of weak absorption lines and hence should be reliable, but is based on
only one data point:  NGC 2024 IRS 2.  If this sightline is abnormal in any way, then the ratio
does not necessarily apply to other lines of sight such as \wforty.

\subsection{Dust Indicators}

The \av\ calculated by Crutcher and Chu (1982) and Smith et al. (1985) assumes a
``normal'' extinction law ($R_V \sim 3$), which may be incorrect. The ratio of
visual to selective extinction, $R_V$, is a grain size distribution indicator (\cite{ccm89}):
$R_V < 3$ implies an abundance of small grains compared to the typical size distribution,
whereas 
$R_V > 3$  implies that larger grains dominate the extinction.  As interstellar clouds
collapse the dust grains tend to agglomerate, eliminating the smaller particles (\cite{j80}). 
If the line of sight toward \wforty\ does indeed sample the edge of a molecular cloud, then
we would expect a population of larger grains and $3 < R_V < 5$.  Smith et al.
(1985) found $(B-V)$ = 2.2 for
\wforty\ and assuming it is an OB star, $E_{B-V} \simeq 2.5$.  For $R_V = 3-5$, we get a range
in \av\ of 7.5 to 12.5, which agrees with the previous work by Crutcher and Chu (1982) and
Smith et al. (1985).

Our calculated value of \av\ can be used to predict the amount of CO expected toward \wforty .
As discussed in Section 2, however, the extinction and CO absorption may not be well-correlated
in the PDR, leading us to slightly {\em overestimate} $N_{CO}$ based on \av.  Conversely,
most studies of the CO/\av\ ratio have not addressed the depletion of CO into ice
mantles (which may be as high as 40 \%, \cite{cetal95}).  Since CO is almost entirely in the
gas-phase for the \wforty\ line of sight, these studies will {\em underestimate} the predicted
CO column density.  Using radio maps, spectroscopic data, and star counts in the Taurus and
$\rho$ Oph clouds, Frerking, Langer, and Wilson (1982) found:
\begin{equation}
N(C^{18}O) = 1.7 \times 10^{14} (A_V - 1.3)
\end{equation}
for $N$ in cm$^{-2}$ and $4 < A_V < 21$.  This relation predicts $N_{CO} =  (0.5 - 1.0) \times
10^{18}$ cm$^{-2}$ (with $N($CO$)/N($C$^{18}$O$) = 490$) for our line of sight, very nearly the
same as we have measured.  Interestingly, it may be that the PDR effect nearly balances the
depletion effect.

There is no strong
correlation between \ctwo\ and $E_{B-V}$ (\cite{vdb89}) though other  dust
indicators have not been investigated.  The lack of correlation may be due to the fact 
that many of the stars included in
\ctwo\ absorption studies to date have been distant supergiants where some of the sightline extinction
is caused by diffuse clouds, which lack abundant \ctwo.  

Comparing the \wforty\ sightline to others in Table 7, it is apparent that CO and
\ctwo\ generally increase with \av\, but the data are too scattered to draw any strong
conclusions.  

The measurement of CO and \ctwo\ absorption toward Cyg OB2
No. 12, the classic diffuse cloud line of sight with
$A_V \simeq 10$, is quite interesting.  Lutz and Crutcher (1983) found $N_{C_2} = (3.0 \pm 0.2)
\times 10^{14}$ cm$^{-2}$ (adjusted for new Phillips system $f$-values in Lambert, Sheffer, and
Federman (1995)), about half of the abundance on the \wforty\ sightline.  Recently, McCall et al.
(1998) measured CO IR absorption toward Cyg OB2 No. 12 yielding $N_{CO} = 2 \times 10^{16}$ cm$^{-2}$,
a factor of 60 less than what we find for \wforty\ over nearly the same total visual extinction.  
The 4.7
$\mu$m CO ice feature is not seen on this line of sight; therefore, depletion of CO onto ice
mantles cannot readily explain the discrepancy.  This clearly shows that the gaseous CO to dust
ratio changes from diffuse to denser environments.

As an interesting side note, we have compared the total hydrogen column density (assuming all of
it to be molecular) for both \wforty\ and NGC 2024 IRS 2 to the
hydrogen/reddening correlations found by Bohlin, Savage, and Drake (1978) and Dickman (1978)
(Figure 5).  The $E_{B-V}$ for NGC 2024 IRS 2 assumes $A_V = 21$ (\cite{letal94}; \cite{jpl84}) 
and $R_V$ ranging from 3 to 5.  
The correlations (``intercloud'' and ``cloud'') from Bohlin,
Savage, and Drake (1978) are based on L$\alpha$ absorption for lightly reddened sightlines with
$E_{B-V} < 0.6$, while the relation from Dickman (1978), derived from CO emission, is good to
$E_{B-V} \sim 3$.  In Figure 5 we have extrapolated out to
$E_{B-V} = 10$.  The data from \wforty\ and NGC 2024 IRS2
agree with the extrapolated relations surprisingly well, within factors of 2 or so.

\section{Conclusions}

We have used \twelveco\ and \ctwo\ IR and visible absorption lines to investigate the line of
sight toward
\wforty.  The \ctwo\ data were obtained at the 4 m Mayall Telescope
at KPNO and the \twelveco\ data from the CSHELL on the IRTF.  The CO ice band at 4.7 \micron\
was not detected.  This sightline is clearly much more dense than diffuse, based on $A_V \sim 9$
and a distance of only 400 pc (we calculate a range in \av\ of 7.5 to 12.5, assuming a
population of large grains).  Our \twelveco\ and \ctwo\ data show:

1. The \ctwo\ excitation conditions, $T \simeq 40$ K and $n \sim 250$ cm$^{-3}$, agree with
limits determined from CO emission and absorption. Since
the \av\ is large and both absorption and emission measures are available, we postulate that the \wforty\
sightline has multiple translucent components.

2.  The non-detection of CO ice indicates a CO depletion of $\delta < 1\%$.

3.  Comparing to other sightlines in Table 7, we find an overall increase in $N_{CO}$ and
$N_{C_2}$ with increasing \av.  The data, however, are too
scattered to draw any further conclusions at this point.  The ratio of CO to \ctwo\ appears to
increase from diffuse to translucent and molecular sightlines probably due to the
self-shielding of CO.

4.  The column density of CO toward \wforty\ is $\sim 60$ times that found for Cyg
OB2 No. 12, the classic diffuse line of sight (\cite{mccall98}), despite very similar \av\ values. 
This is not a depletion effect, and suggests that the CO-to-dust ratio changes from diffuse to
dense environments.

5.  Finally, the relationship of hydrogen column density to interstellar reddening
(\cite{bsd78}; \cite{dickman78}) was found to be roughly consistent with recent data out to
$E_{B-V}
\sim 10$ (Figure 5). Contrary to the CO-to-dust ratio, the hydrogen-to-dust ratio appears to be
valid from diffuse to dense regimes.

\acknowledgments

This research was supported by NASA Graduate Student Research Program grant (NGT5-50032) to R.
Y. Shuping and NASA grant NAG5-4184 to T. P. Snow.  We would like to thank the staff and
operators at KPNO, the IRTF (J. Rayner), and the UKIRT (T. Kerr) for their help and tireless
duty through the night.  Many thanks to J. Black and the referee S. Federman, whose comments and
suggestions greatly improved the content and quality of this paper. B. L. Lutz would
like to thank K. Sheth who carried out the curve-of-growth analysis and temperature fits for the
\ctwo\ lines.   R. Y. Shuping would like to thank B-G Andersson, J. H. Lacy, and D.
Jansen for their helpful input, and also D. E. Schutz for useful conversations and support.

%\clearpage

% Now comes the reference list.  In this document, we used \cite to call
% out citations, so we must use \bibitem in the reference list, which
% means we use the LaTeX thebibliography environment.  Please note that
% \begin{thebibliography} is followed by a null argument.  If you forget
% this, mayhem ensues, and LaTeX will say "Perhaps a missing item?" when
% you run it.  Do not call us, do not send mail when this happens.  Put
% the silly {} after the \begin{thebibliography}.
%
% Each reference has a \bibitem command to define the citation format
% to be placed in the text (in []) and the symbolic tag used for 
% cross referencing (in {}).
%
% See sample1.tex, or the AASTeX guide, for an alternative to the \cite-
% \bibitem command.

% We start with a table using the "deluxetable" environment.
%
% The caption contains only the caption text.  The "Table N." identification
% is generated by the \tablecaption command on its own.  It is necessary to 
% \label tables and figures *after* the caption has been specified because 
% the table/figure number is generated by the caption, not by \begin{whatever}.
% The column headings are specified within a \colhead command and all the
% column headings are included within a single \tablehead command.  The
% \enddata command comes at the end of the data, and the table is closed with 
% an \end{deluxetable} command.  It the table is too wide for the page, \small
% (11pt), \footnotesize (10pt), or \scriptsize (8pt) may be used inside
% the deluxetable environment - the table will still be double-spaced.  For
% even wider tables see the AASTeX guide.

\clearpage

%Table 1

\begin{deluxetable}{ccc}
\tablecaption{Observed \ctwo\ Phillips System Lines\label{tbl1}}
\tablewidth{400pt}
\tablehead{
\colhead{Transition} & \colhead{Rest Wavelength (\AA)}   &
\colhead{Equivalent Width (m\AA)} 
} 
\startdata
R0        &  8757.7  &  $37.0 \pm 3.0$  \nl
R2        &  8753.9  &  $52.5 \pm 1.5$\tablenotemark{a}  \nl
Q2        &  8761.2  &  $55.0 \pm 5.0$  \nl
P2        &  8766.0  &  $18.0 \pm 3.0$  \nl
R4        &  8751.6  &  $42.5 \pm 7.5$\tablenotemark{b}  \nl
Q4        &  8763.7  &  $36.0 \pm 3.0$  \nl
P4 + Q8   &  8773.3  &  $43.0 \pm 4.0$  \nl
R6        &  8750.8  &  $22.8 \pm 3.0$  \nl
Q6        &  8767.7  &  $31.5 \pm 4.5$  \nl
P6        &  8782.3  &  $12.5 \pm 2.5$  \nl
P8        &  8792.6  &  $9.5 \pm 2.0$   \nl
Q10       &  8780.1  &  $17.5 \pm 2.5$  \nl
Q12       &  8788.5  &  $6.5 \pm 2.0$   \nl
\enddata
\tablenotetext{a}{R2 and R10 are blended}
\tablenotetext{b}{R4 and R8 are blended}
\end{deluxetable}
%--------------------------------------------

%Table 2

\begin{deluxetable}{ccc}
\tablecaption{\ctwo\ Column Densities\label{tbl2}}
\tablewidth{350pt}
\tablehead{
\colhead{$J$} & \colhead{$E_J/k$ (K)}   &
\colhead{$N_J\ (\times 10^{13}$ cm$^{-2}$)} 
} 
\startdata
0   &  0.000     &  $5.4 \pm 0.8$   \nl
2   &  15.635    &  $19.8 \pm 2.0$  \nl
4   &  52.114    &  $11.2 \pm 1.6$  \nl
6   &  109.430   &  $9.2 \pm 0.7$  \nl
8   &  187.572   &  $5.6 \pm 0.7$  \nl
10  &  286.526   &  $4.1 \pm 0.4$  \nl
\hline
Total\tablenotemark{a} &          &  $70 \pm 4$     \nl
\enddata
\tablenotetext{a}{Based on Boltzmann Analysis for both 39 and 126 K}
\end{deluxetable}
%------------------------------------------------------------

%Table 3
 
\begin{deluxetable}{ccccc}
\footnotesize
\tablecaption{CSHELL \twelveco\ Observations Summary\label{tbl3}}
\tablewidth{0pt}
\tablehead{
\colhead{Wavelength Range ($\mu$m)} & \colhead{\twelveco\ Lines}   &
\colhead{Int. Time (min)}   &
\colhead{S/N\tablenotemark{a}} & 
\colhead{Standard Stars}
} 
\startdata
2.334-2.340 & R3, R4 and R5 & 25  & 90 & BS 6714, 7110 and 5 Aql \nl
2.340-2.346 & R0,1, and 2 & 25 & 90 & Spica \nl
\enddata
\tablenotetext{a}{S/N for the regions surrounding the applicable lines.}
\end{deluxetable}
%------------------------------------------------------------

%Table 4

\begin{deluxetable}{cccc}
\tablecaption{Observed \twelveco\ Lines ($v = 2 - 0$)\label{tbl4}}
\tablewidth{0pt}
\tablehead{
\colhead{Transition} & \colhead{Rest Wavelength ($\mu$m)\tablenotemark{a}}   &
\colhead{$f$-value ($\times 10^{-8}$)\tablenotemark{b}}   &
\colhead{Equiv. Width (m\AA)}
} 
\startdata
R0 & 2.34530523 & 8.78 & $130 \pm 20$  \nl
R1 & 2.34326929 & 5.89 & $100 \pm 10$  \nl
R2 & 2.34127497 & 5.33 & $60 \pm 17$  \nl
R3 & 2.33932336 & 5.11 & $< 20$  \nl
R4 & 2.33741273 & 5.00 & $< 20$  \nl
R5 & 2.33554435 & 4.94 & $< 20$  \nl
\enddata
\tablenotetext{a}{\cite{mm74}}
\tablenotetext{b}{\cite{ct83}}
\end{deluxetable}
%------------------------------------------------------------

%Table 5

\begin{deluxetable}{ccc}
\tablecaption{\twelveco\ Column Densities}
\tablewidth{0pt}
\tablehead{
\colhead{$J$} & \colhead{$E_J/k$ (K)}   & 
\colhead{$N_J\ (\times 10^{17}$ cm$^{-2})$}
} 
\startdata
0 & 0     &  $ 5.0 \pm 2.0 $ \nl
1 & 5.532 &  $ 4.0 \pm 1.5 $ \nl
2 & 16.60 &  $2.2 \pm 0.7 $ \nl
3 & 33.19 &  $< 0.8$ \nl
4 & 55.32 &  $< 0.8$ \nl
5 & 82.98 &  $< 0.8$ \nl
\hline
Total &   & $11 \pm 2$   \nl
\enddata
\end{deluxetable}

%-----------------------------------------

%Table 6

\begin{deluxetable}{ccccccc}
\footnotesize
\tablecaption{\wforty\ Line of Sight Physical Properties\label{tbl6}}
\tablewidth{0pt}
\tablehead{
\colhead{Species} & \colhead{$V_{LSR}$ (km s$^{-1})$} & \colhead{$b$ (km s$^{-1}$)} &
\colhead{$N$ (cm$^{-2})$}   &
\colhead{$T_{ex}$ (K)} & \colhead{$n_H$\tablenotemark{c} (cm$^{-3}$)}  &
\colhead{$\log{(nT)}$}
} 
\startdata
\ctwo     & ...        & 1.25 & $7.0 \pm 0.4 \times 10^{14}$ & 39 and 126\tablenotemark{b}    &
$\sim 250$  &  ... \nl
$^{13}$CO Em.\tablenotemark{a} &   8     & ... & $1.2 \times 10^{18}$  &  ...  &  ... &
$\sim 4$ 
\nl
$^{12}$CO & $2 \pm 2$ & ... &$1.1 \pm 0.2 \times 10^{18}$   & $> 7$ & $< 450$\tablenotemark{d} &
...
\nl
\enddata
\tablenotetext{a}{\cite{cc82}}
\tablenotetext{b}{$T = 126$ K component due to radiative pumping.}
\tablenotetext{c}{H $+$ H$_2$}
\tablenotetext{d}{Assuming $T \simeq 40$ K, see Section 4.1}
\end{deluxetable}
%------------------------------------------------------------

%Table 7

\begin{deluxetable}{ccccccc}
\footnotesize
\tablecaption{Sightlines With CO and \ctwo\ Observations}
\tablewidth{0pt}
\tablehead{
\colhead{Sightline} & \colhead{$N_{CO}$ (cm$^{-2}$)\tablenotemark{a}} &
\colhead{$N_{C_2}$ (cm$^{-2}$)\tablenotemark{b}} & \colhead{CO/\ctwo} &
\colhead{Cloud Type\tablenotemark{c}} & \colhead{\av}   &
\colhead{References\tablenotemark{d}}
}
\startdata
W40 IRS 1a  &	$1.1 \pm0.2$(18)  & $7.0 \pm 0.4$(14)  &  $1600 \pm 600$  & T	& 7.5 -- 12.5	& This Work   \nl  
Cyg OB2 \#12 &	2(16)	& $3.0 \pm 0.2$(14)	& $67 \pm 5$	& D	& $\sim 10$	& 1, 2  \nl 
HD94413	& 0.7--1.2(16)\tablenotemark{e} &	2.8(13)  &  250--430  &	T	  &  2.4	&  3  \nl  
HD154368	& 0.6--1.5(16)\tablenotemark{e}  &	4.6(13)	  &  130--330	&  T	&  2.5 &  3  \nl  
HD169454	& 0.55--1.8(16)\tablenotemark{e}  &	5.6(13)  &	100--320	&  T	&  3.3	& 3  \nl  
$o$ Per	& 1.1(15)	&  1.8(13)	& 60	&  D	&  0.9	&  4 \nl  
HD27778	& 2.5(16)  &3.0(13)	&  830	&  T &  1.2	&  4  \nl
$\rho$ Oph A	& 1.9(15)	&  2.1(13)	&  90	&  T	&  1.4	&  4,7  \nl
$\zeta$ Oph	& 2.3(15)	&  $1.79 \pm 0.06$(13)	&  130	&  D/T	&  0.96	&  4  \nl 
20 Aql	& 3(15)	&  4.2(13)	&  71	&  D/T	&  0.99	&  4  \nl 
HD207198	& 2.6(15)	&  2.4(13)	&  110	&  T	&  1.9	&  4  \nl
HD21483      &  1.0(18)  &  7.4(13)    &  100 - $10^5$ & T & 1.7 & 4  \nl
$\zeta$ Per  &  1.2(15)  &  2.8(13)    &  43           & D/T & 1.0 & 4  \nl
X Per        &  5.0(15)  &  4.2(13)    &  120          & T & 1.4 & 4  \nl
HD 26571     &  6.0(16)  &  8.8(13)    &  680          & D/T & 0.9 & 4  \nl
AE Aur       &  1.3(15)  &  4.6(13)    &  28           & T & 1.6 & 4  \nl
HD110432     &  1.0(15)  &  2.4(13)    &  42           & T & 1.2 & 4  \nl
$\pi$ Sco    &  1.0(12)  &  $<1.0(12)$ & $>1$           & D & 0.19 & 5  \nl
$\beta^1$ Sco & 1.2(13)  &  $<1.0(12)$ & $>12$           & D & 0.62 & 5  \nl
$\omega^1$ Sco &4.0(13)  &  $<2.0(12)$ & $>23$           & D & 0.68 & 5  \nl
$\chi$ Oph  &  3.0(14)  &  2.8(13)    &  11           & T & 1.2 & 4  \nl
9 Cep        &  1.7(13)  &  1.0(13)    &  2           & D\tablenotemark{f} &  1.5 & 4  \nl
HD 210121    &  3.0(15)\tablenotemark{e} & 5.2(13) & 58 & T (High Lat.) & $\sim 1$ & 6  \nl
$\lambda$ Cep & 1.4(15)  &  1.4(13)    &  100            & D\tablenotemark{f} & 1.8 & 4  \nl
\enddata
\tablecomments{$X(Y) = X \times 10^{Y}$}
\tablenotetext{a}{All column densities derived from UV or IR absorption lines except where
noted.}
\tablenotetext{b}{Where appropriate, the column density for \ctwo\ has been adjusted to account
for the new $f$-value derived by Lambert, Sheffer, and Federman (1995).}
\tablenotetext{c}{Cloud Types:  D -- Diffuse, T -- Translucent}
\tablenotetext{d}{References:  1 -- \cite{mccall98}, 2 -- \cite{lc83}, 3 -- \cite{gvdb94}, 4 -- \cite{fetal94} and 
references therein, 5 -- \cite{lsf95}, 6 -- \cite{getal92}, 7 -- \cite{fetal99}}
\tablenotetext{e}{From mm emission lines}
\tablenotetext{f}{Possibly more than one cloud.}
\end{deluxetable}
%------------------------------------------------------------

\clearpage

\begin{figure}
\plotone{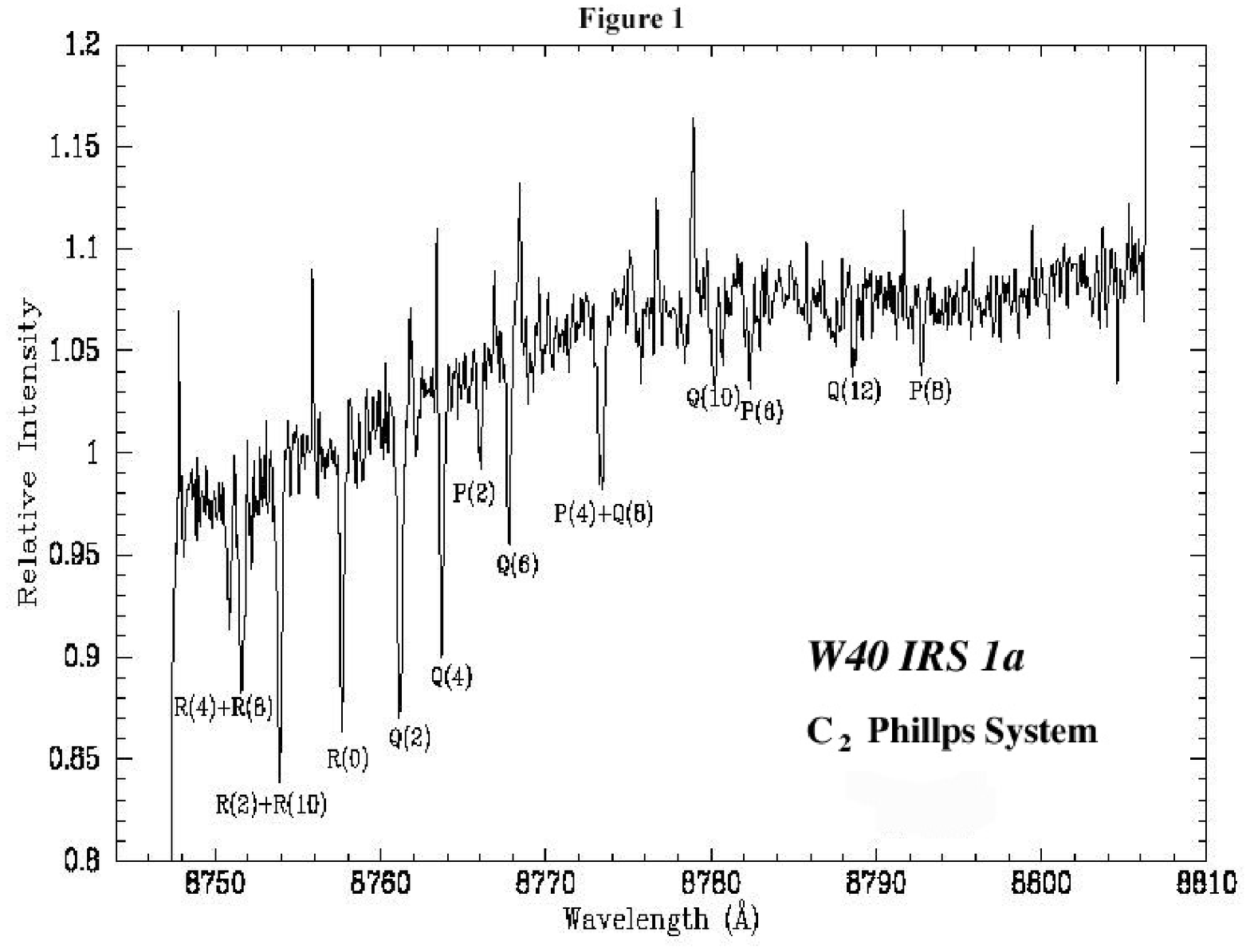}
\caption{\ctwo\ Phillips system toward \wforty.}
\end{figure}

\begin{figure}
\plotone{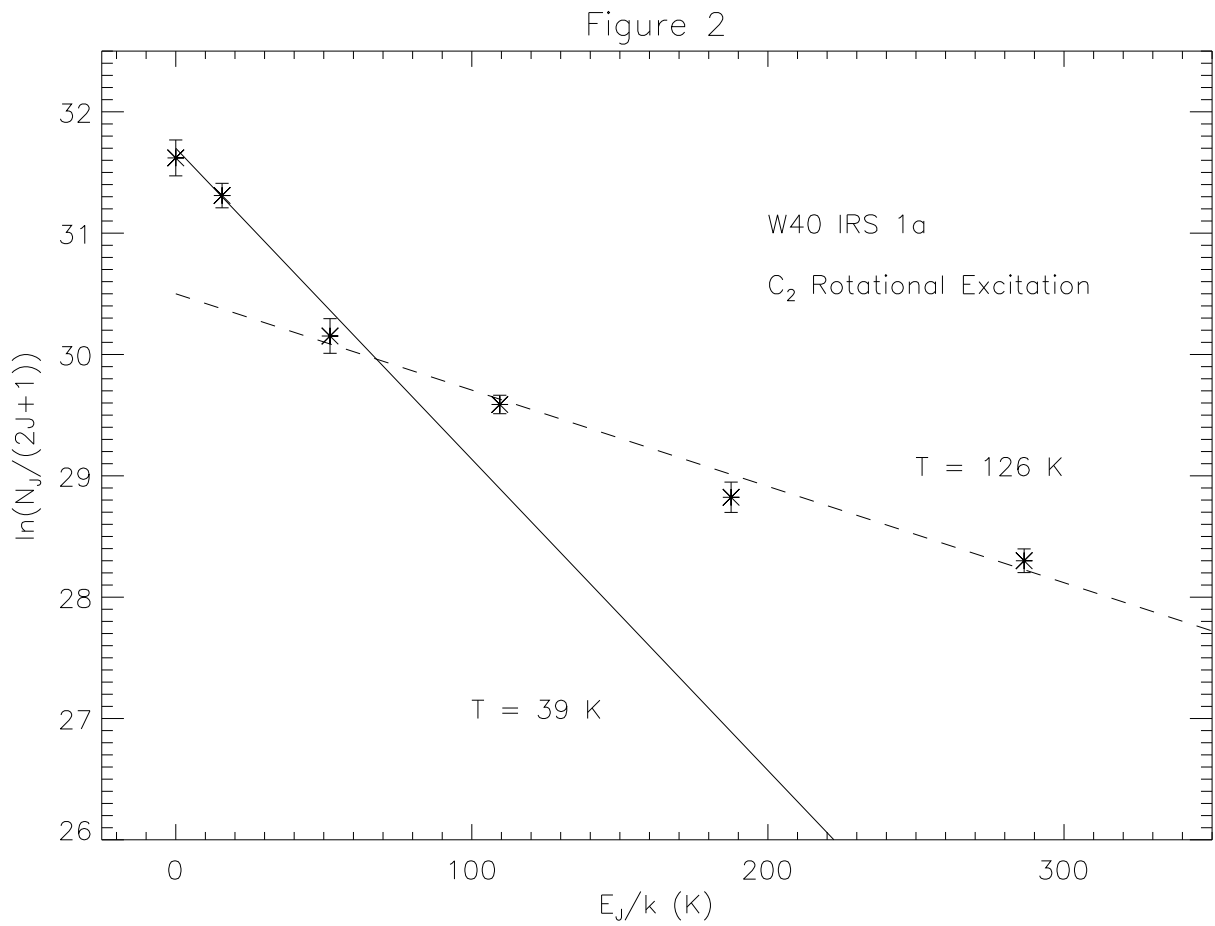}
\caption{Boltzmann plot for rotational levels of \ctwo:  Two-temperature model, 39 K (solid)
and 126 K (dashed).}
\end{figure}

\begin{figure}
\plotone{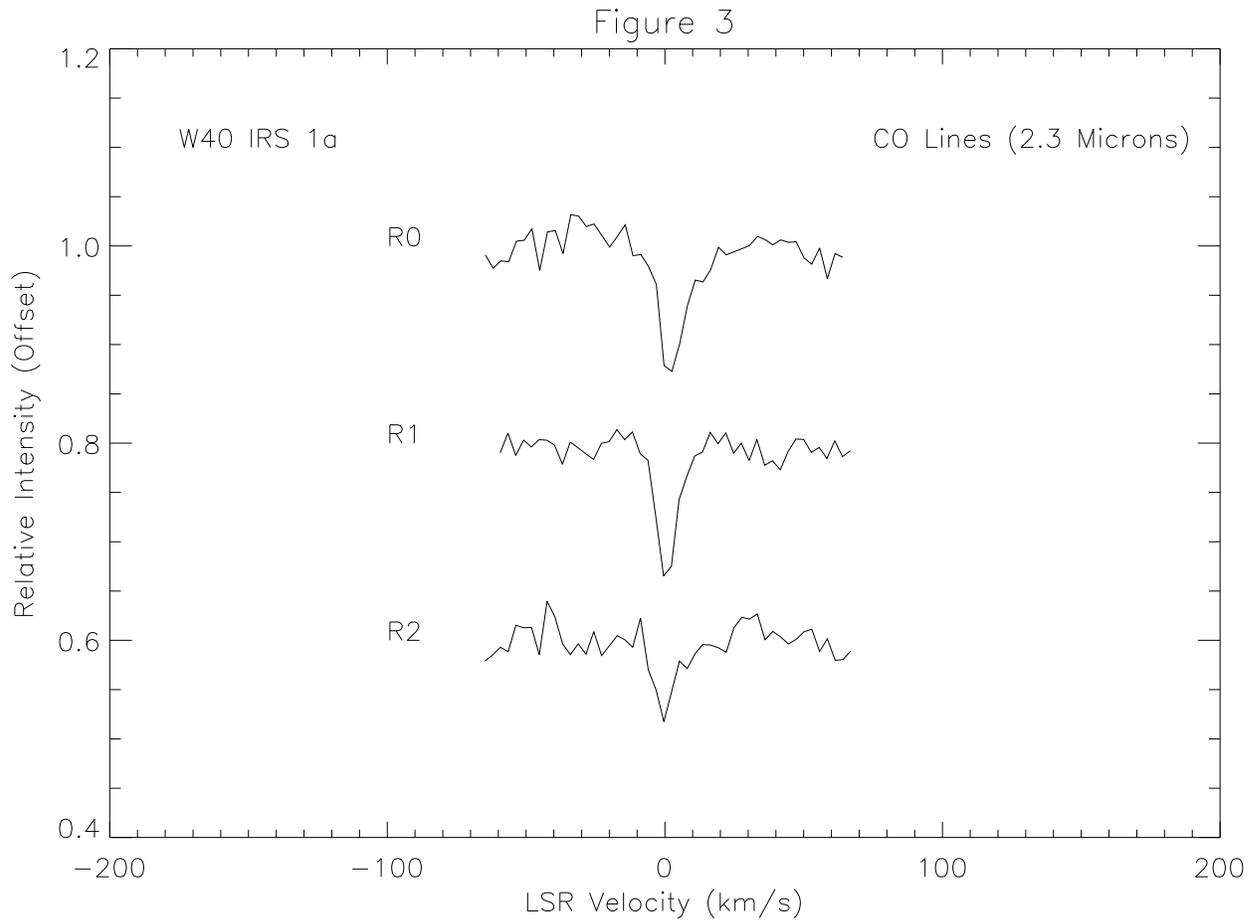}
\caption{\twelveco\ lines at 2.3 $\mu$m toward \wforty.  The continuum for R1 and R2 is
offset for clarity.}
\end{figure}

\begin{figure}
\plotone{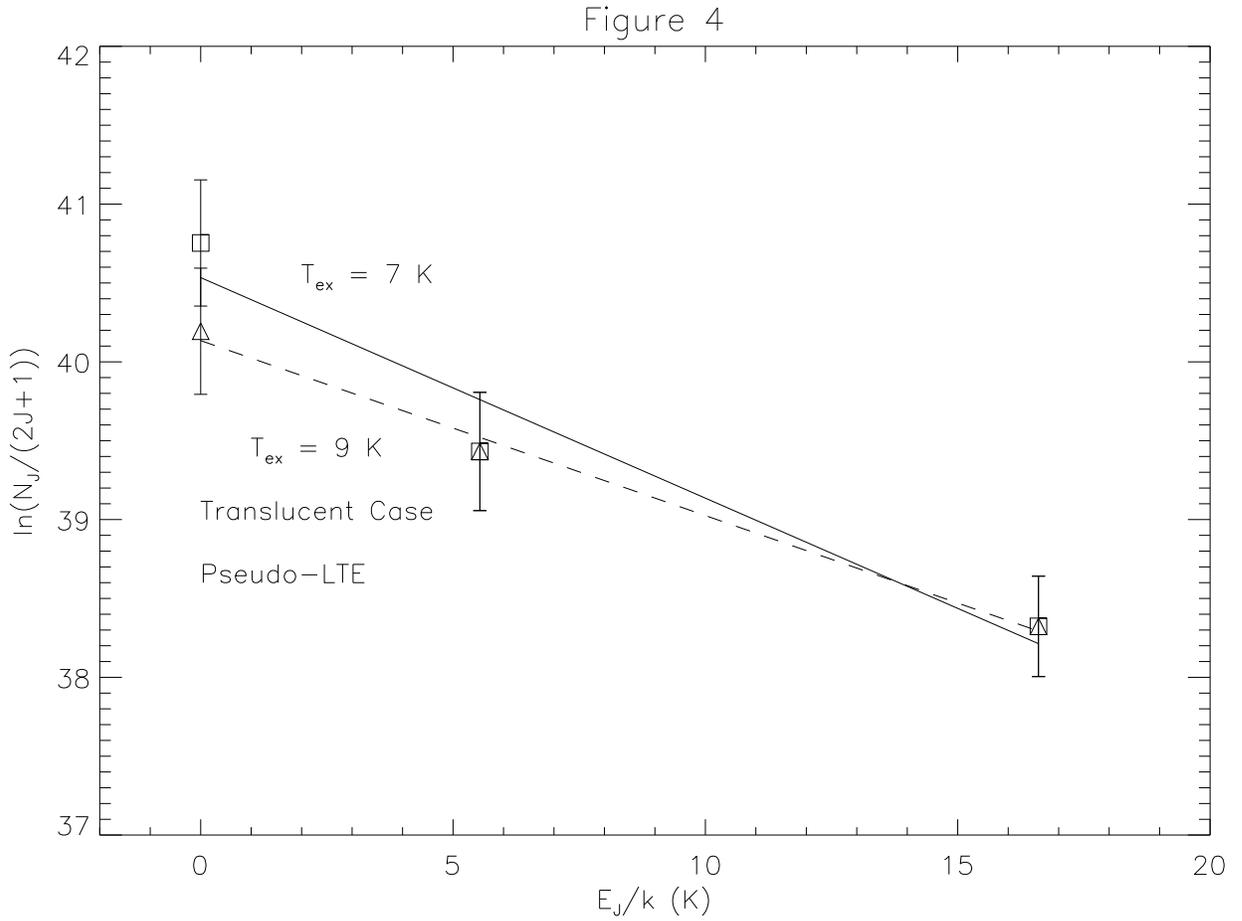}
\caption{Boltzmann plot for rotational levels of \twelveco.  Squares are observed column
densities from Table 5 with $T_{ex} = 7$ K (solid line).  Triangles are pseudo-LTE populations
assuming material on the line of sight is translucent in nature with temperature fit ($T_{ex} =
9$ K,dashed line).  See section 4.1 for discussion.}
\end{figure}

\begin{figure}
\plotone{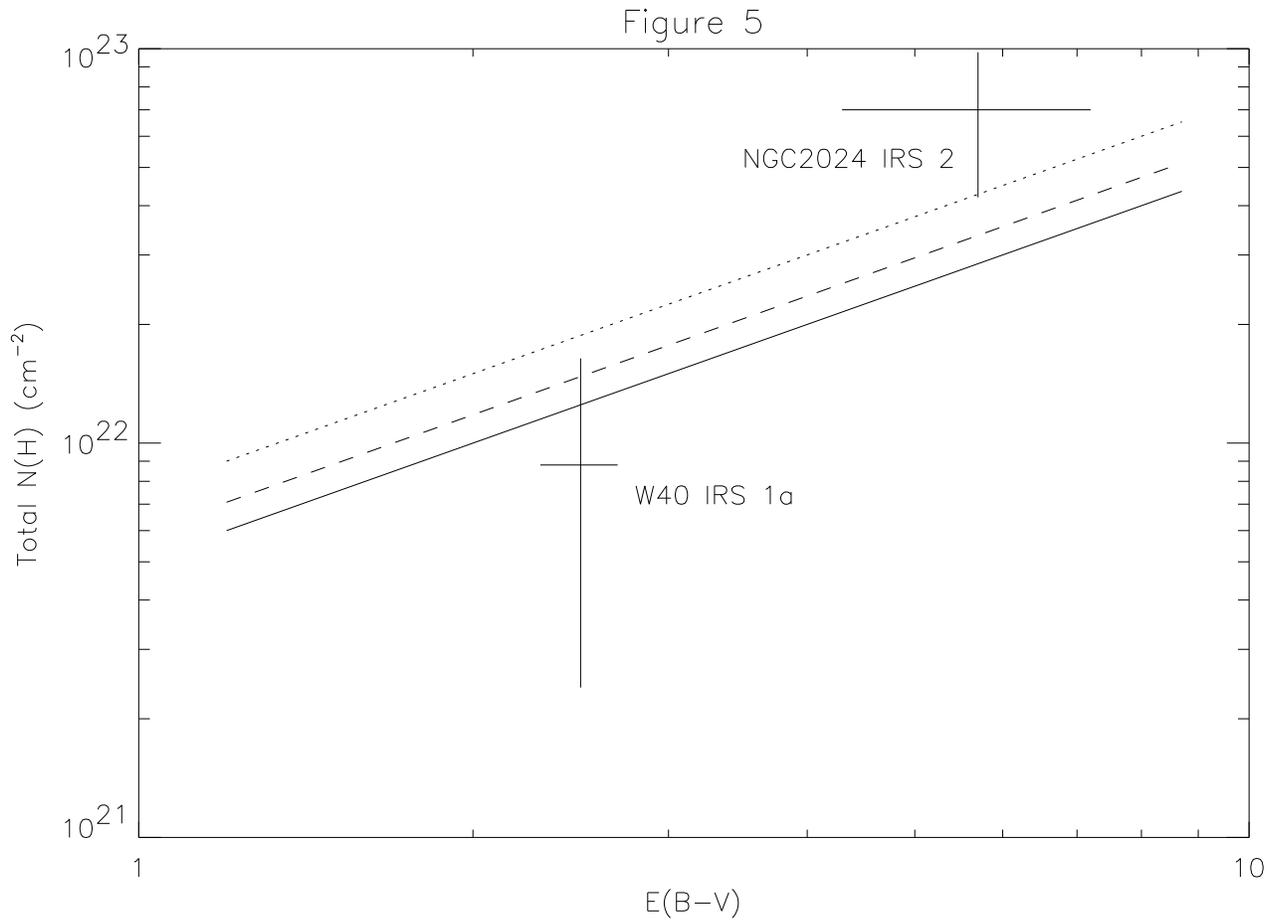}
\caption{Correlation of reddening and total hydrogen column density.  Solid line is
``intercloud'' relationship extrapolation from Bohlin, Savage, and Drake (1978); dashed line is
the ``cloud'' extrapolation; dotted line is relationship found by Dickman (1978).  The total H
column density for both
\wforty\ and NGC 2024 IRS 2 assumes all hydrogen is molecular.  See Section 4.3 for further
discussion.}
\end{figure}

%\figcaption{\ctwo\ Phillips system toward \wforty.}
%
%\figcaption{Boltzmann plot for rotational levels of \ctwo:  Two-temperature model, 39 K (solid)
%and 126 K (dashed).}
%
%\figcaption{\twelveco\ lines at 2.3 $\mu$m toward \wforty.  The continuum for R1 and R2 is
%offset for clarity.}
%
%\figcaption{Boltzmann plot for rotational levels of \twelveco.  Squares are observed column
%densities from Table 5 with $T_{ex} = 7$ K (solid line).  Triangles are pseudo-LTE populations
%assuming material on the line of sight is translucent in nature with temperature fit ($T_{ex} =
%9$ K,dashed line).  See section 4.1 for discussion.}
%
%\figcaption{Correlation of reddening and total hydrogen column density.  Solid line is
%``intercloud'' relationship extrapolation from Bohlin, Savage, and Drake (1978); dashed line is
%the ``cloud'' extrapolation; dotted line is relationship found by Dickman (1978).  The total H
%column density for both
%\wforty\ and NGC 2024 IRS 2 assumes all hydrogen is molecular.  See Section 4.3 for further
%discussion.}

\end{document}